\documentstyle[aps,prl,epsf,twocolumn]{revtex}
\begin{document}
\draft
\title{{\rm Physical Review Letters}\hfill {\sl Version of \today}\\~~\\
The Geometry of Developing Flame Fronts: Analysis with Pole Decomposition}
\author {Oleg Kupervasser\cite{oleg} Zeev Olami\cite{zeev}  and
Itamar  Procaccia\cite{procaccia}  }
\address{Department of~~Chemical Physics,\\
 The Weizmann Institute of Science,
Rehovot 76100, Israel} \maketitle
\narrowtext
\begin{abstract}
The roughening of expanding flame fronts by the accretion of cusp-like
singularities
is a fascinating example of the interplay between instability, noise and
nonlinear
dynamics that is reminiscent of self-fractalization in Laplacian growth
patterns.
The nonlinear integro-differential equation
that describes the dynamics of expanding flame fronts is amenable to analytic
investigations using pole decomposition. This powerful technique
allows the development of a satisfactory understanding of the qualitative and
some quantitative aspects of the complex geometry that develops in
expanding flame fronts.
\end{abstract}
\narrowtext
The study of growing fronts in nonlinear physics \cite{Pel} offers
fascinating examples of
spontaneous generation of fractal geometry\cite{BS,Vic}. Advancing fronts
rarely remain
flat;
usually they form either fractal objects with contorted and ramified
appearance, like
Laplacian growth patterns and diffusion limited agregates (DLA)\cite{81WS},
or they remain
graphs,
but they ``roughen" in the sense of producing self-affine fractals whose
``width" diverges
with the linear scale of the system with some characteristic exponent.
The study
of interface growth where the roughening is caused by the noisy environment,
with either annealed or quenched noise, was
a subject of active research in recent years\cite{93Mea,95HHZ}. These
studies met
considerable success
and there is significant analytic understanding of the nature of the
universality classes
that can be expected. The study of interface roughening in system in which
the flat surface
is inherently unstable
is less developed. One interesting example that attracted attention is
the Kuramoto-Sivashinsky equation \cite{78Kur,77Siv} which is known to
roughen in 1+1
dimensions but is
claimed not to roughen in higher dimensions\cite{92LP}. Another outstanding
example is
Laplacian growth patterns\cite{84SB}. This Letter is motivated by a new example
of the dynamics of outward propagating flames whose front wrinkles and
fractalizes
\cite{94FSF}. We will see that
this problem has many features that closely resemble Laplacian growth, including
the existence of a single finger in channel growth versus tip splitting in
cylindrical
outward growth, extreme sensitivity to noise, etc. In the case of flame
fronts the
equation of motion is amenable to analytic solutions and as a result we
can understand some of these issues.

The physical problem that motivates this analysis is that of pre-mixed
flames which
exist as self-sustaining fronts of exothermic chemical reactions in gaseous
combustion. It had been known for some time that such flames are
intrinsically unstable
\cite{44Lan}.
It was reported that such flames develop characteristic structures which
includes cusps,
and that under usual experimental conditions the flame front accelerates as
time goes
on \cite{89GIS}. In recent work Filyand et al. \cite{94FSF} proposed an
equation of motion
that
is motivated by the physics and seems to capture a number of the essential
features
of the observations.  The equation is written in cylindrical geometry and is for
$R(\theta,t)$ which is the modulus of the radius vector on the flame front:
\begin{eqnarray}
{\partial R \over \partial t}&=&
 {U_b\over 2{R_0}^2(t)}\left({\partial R \over \partial \theta }\right)^2
 +{D_M\over {R_0}^2(t)}{\partial^2 R\over \partial\theta^2}\\ \nonumber &+&
{\gamma
 U_b\over 2R_0 (t)} I(R)+U_b \ . \label{Eqdim}
\end{eqnarray}
Here 0$ <\theta <  2\pi$ is an angle and the constants $U_b,D_M$ and
$\gamma$ are the
front velocity for an ideal cylindrical front, the Markstein diffusivity
and the thermal
expansion coefficient respectively. $R_0(t)$ is the mean
radius of the propagating flame:
\begin{equation}
R_0 (t)={1\over 2\pi}\int_{0}^{2\pi}R(\theta,t)d\theta \  . \label{R0}
\end{equation}
The functional $I(R)$ is best represented in terms
of its Fourier decomposition. Its Fourier component is $|k|R_k$ where $R_k$
is the
Fourier component of $R$.

Numerical simulations of the type reported in ref.\cite{94FSF} are presented in
Fig.1. The flame front $R(\theta,t)$ is shown at four equal time intervals.
The two
most prominent features of these simulations are the wrinkled multi-cusp
appearance
of the fronts and its acceleration as time progresses. One observes the
phenomenon
of tip splitting in which new cusps are added to the
growing fronts between existing cusps. Both experiments and simulations
indicate that for large times $R_0$ grows as a power in time
\begin{equation}
R_0(t) = (const+t)^{\beta} \ , \label{accel}
\end{equation}
with $\beta>1$, (of the order of $1.5$) and that the width of the interface $W$
grows with $R_0$ as
\begin{equation}
W(t) \sim R_0(t)^\chi \ , \label{scaling}
\end{equation}
with $\chi<1$ (of the order of 2/3). The understanding of these two
features and
the derivation of the
scaling relation between $\beta$ and $\chi$ are the main aims of this Letter.

Equation (\ref{Eqdim}) can be written as a one-parameter equation by
rescaling $R$ and $t$
according to $r\equiv RU_b/ D_M$,  $\tau\equiv tU_b^2/ D_M$.
Computing the derivative of Eq.(\ref{Eqdim}) with respect to $\theta$ and
substituting
the dimensionless variables one obtains:
\begin{equation}
{\partial u \over \partial \tau}={u \over r_{0}^2}
   {\partial u \over \partial \theta}+{1\over r_{0}^2}{\partial^2u
   \over \partial \theta^2}+{\gamma\over 2 r_0}I\{ u\} \ . \label{eqfinal}
\end{equation}
where $u\equiv {\partial r \over \partial \theta}$.
To complete this equation we need a second one for $r_0(t)$, which is obtained
by averaging (\ref{Eqdim}) over the angles and rescaling as above. The result is
\begin{equation}
{dr_0 \over d\tau}={1  \over 2r_{0}^2}{1 \over
2\pi}\int_{0}^{2\pi}u^2d\theta +1 \ .
\label{eqr0}
\end{equation}
These two equations are the basis for further analysis

Following \cite{82LC,85TFH,90J,84BF,89J,95J} we expand now the solutions
$u(\theta,\tau)$
in
poles whose position $z_j(\tau)\equiv x_j(\tau)+iy_j(\tau)$ in the complex
plane
is time dependent:
\begin{eqnarray}
u(\theta,\tau)&=&\sum_{j=1}^{N}\cot \left[{\theta-z_j(\tau) \over 2}\right]
   + c.c.  \label{upoles} \\
&=&\sum_{j=1}^{N}{2\sin [\theta-x_j(\tau)]\over
\cosh [y_j(\tau)]-\cos [\theta-x_j(\tau)]}\ , \nonumber
\end{eqnarray}
\begin{equation}
r(\theta,\tau)=2\sum_{j=1}^{N}{\ln \Big[\cosh (y_j(\tau))-\cos
(\theta-x_j(\tau))
\Big]}+C(\tau) \ . \label{rpoles}
\end{equation}
In (\ref{rpoles}) $C(\tau)$ is a function of time. The function (\ref{rpoles})
is a superposition of quasi-cusps (i.e. cusps that are rounded at the tip). The
real part of the pole position (i.e. $x_j$) describes the angle coordinate
of the maximum
of the quasi-cusp, and the imaginary part of the pole position (i.e $y_j$)
is related
the height of the quasi-cusp. As $y_j$ decreases (increases) the height of
the cusp
increases (decreases). The physical motivation for this
representation of the solutions should be evident from Fig.1.

The main advantage of this representation is that the propagation and
wrinkling of the
front can be described now via the dynamics of the poles and of $r_0(t)$.
Substituting
(\ref{upoles}) in (\ref{eqfinal}) we derive the following ordinary
differential equations for the positions of the poles:
\begin{equation}
- r_0^2{dz_{j}\over d\tau}=\sum_{k=1
  ,k\neq j}^{2N }\cot \left({z_j-z_k\over 2}\right)
  +i{\gamma r_0 \over 2 }sign [Im(z_j)] \ . \label{eqsz}
\end{equation}

After substitution of (\ref{upoles}) in (\ref{eqr0}) we get, using (\ref{eqsz})
the ordinary differential equation for $r_0$,
\begin{equation}
{dr_0\over d\tau}=2\sum_{k=1}^N {dy_k\over d\tau}+2\left( {\gamma\over 2}{N\over
     r_0}-{N^2\over r_0^2}\right) +1 \ . \label{r0pole}
 \end{equation}

In the case of flame fronts propagating in channels of width $L$
ref.\cite{85TFH}
presented a rather complete analysis of the available stationary solutions.
Some aspects
of this analysis are important also for our case of cylindrical geometry,
and we therefore
briefly summarize the main results of \cite{85TFH}. These are:
(i) In noiseless conditions the total number of poles $N_T$ is conserved by the
dynamics. This is also the case in the present problem. (ii) There is only one
stable stationary solution which is geometrically represented by a giant
cusp (or
equivalently one finger) and
analytically by $N(L)$ poles which are aligned on one line parallel to the
imaginary
axis. (iii) The reason for this behaviour is the existence of an attraction
between the
poles
along the real line, and the resulting dynamics merges all the $x$
positions. The $y$
positions are distinct, and the poles are sitting above each others in
positions
$y_{j-1}<y_j<y_{j+1}$ with the maximal $y_{N(L)}$. (iv) If one adds an
additional pole to
such
a solution, this pole (or another) will be pushed to infinity along the
imaginary axis.
If the system has less than $N(L)$ poles it is unstable to the addition of
poles,
and any noise will drive the system towards this unique state. The number
$N(L)$ is
\begin{equation}
N(L)= \Big[{1 \over 2}\left( {L \over 2\pi\nu }+1\right) \Big]\ , \label{NofL}
\end{equation}
where $\Big[ \dots \Big]$ is the integer part, and in the (different)
parametrization of ref.\cite{85TFH} $\nu$ is the coefficient of the viscous
term.
(v) The height of the cusp is proportional to $L$.
we will refer to the solution with these properties as the Thual-Frisch-Henon
(TFH)-cusp solution.

In our problem the outward growth introduces important modifications to the
channel results.
The number of poles in a stable configuration is proportional here to the
radius $r_0$
instead of $L$, but the former
grows in time. The system becomes therefore unstable to the addition of new
poles. If there
is noise in the system that can generate new poles, they will not be pushed
toward infinite
$y$. It is important to stress that any infinitesimal noise (either
numerical or experimental)
is sufficient to generate new poles. These new poles do not necessarily
merge their
$x$-positions with existing cusps.
Even though there is attraction along the real axis as in the channel case,
there is a
stretching of the distance between the poles due to the
radial growth. This may counterbalance the attraction. Our first new idea
is that these two opposing
tendencies define a typical scale denoted as $\cal L$. if we have a cusp
that is made
from the $x$-merging of $N_c$ poles on the line $x=x_c$ and we want to know
whether a
$x$-nearby pole with real coordinate $x_1$ will
merge with this large cusp, the answer depends on the distance $D=r_0|x_c-x_1|$.
There is a length ${\cal L}(N_c,r_0)$ such that if $D>{\cal L}(N_c,r_0)$
then the
single cusp will never merge with the larger cusp. In the opposite limit
the single
cusp will move towards the large cusp until their $x$- position merges and the
large cusp will have $N_c+1$ poles.

This finding stems directly from the equations of motion of the $N_c$
$x$-merged poles and
the single pole at $x_1$. First note that from Eq.\ref{accel} (which is not
explained yet)
it follows that asymptotically $r_0(\tau)=(a+\tau)^\beta$
where $r_0(0)=a^\beta$. Next start from
\ref{eqsz} and write equations for the angular distance $x=x_1-x_c$. It
follows that for
any configuration $y_j$ along the imaginary axis
\begin{equation}
{dx\over d\tau}\le -{2N_c\sin {x} [ 1-\cos{x}]^{-1} \over
   (a+\tau)^{2\beta}}=-{2N_c\cot({x\over 2}) \over (a+\tau)^{2\beta}}\ .
\label{N+1}
 \end{equation}
For small x we get
\begin{equation}
{dx\over d\tau}\le -{4N_c\over x(a+\tau)^{2\beta}}\ .\label{Dpt}
\end{equation}
The solution of this equation is
\begin{equation}
x(0)^2-x(\tau)^2\ge {8N_c \over 2 \beta
-1}(a^{1-2\beta}-(a+\tau)^{1-2\beta})\ .
\label{xtau}
\end{equation}
To find ${\cal L}$ we set $x(\infty) > 0$ from which we find that
the angular distance will remain finite as long as
\begin{equation}
x(0)^2 > {8N_c \over 2 \beta -1}a^{1-2\beta}\ . \label{x0}
\end{equation}
Since $r_0 \sim a^\beta$ we find the threshold angle $x^*$
\begin{equation}
x^* \sim \sqrt N_c r_0^{{(1 - 2\beta) \over 2\beta}} \ , \label{x*2}
\end{equation}
above which there is no merging between the giant cusp and the isolated pole.
To find the actual distance ${\cal L}(N_c,r_0)$ we multiply the angular
distance
by $r_0$ and find
\begin{equation}
{\cal L}(N_c,r_0) \equiv r_0 x^*  \sim \sqrt{N_c} r_0^{{1\over 2\beta}} \ .
\label{ratt}
\end{equation}

To understand the geometric meaning of this result we recall the features
of the
TFH cusp solution. Having a typical length $L$ the number of poles in the
cusp is linear
in $L$. Similarly, if we have in this problem two cusps a distance $2 \cal
L$ apart,
the number $N_c$ in each of them will be of the order of $\cal L$. From
(\ref{ratt})
it follows that
\begin{equation}
{\cal L} \sim  r_0^{{1\over \beta}} \ . \label{scaling1}
\end{equation}
For $\beta>1$ the circumference grows faster than $\cal L$, and therefore
at some
points in time poles that appear between two large cusps would not be
attracted toward
either, and new cusps will appear. We will show later that the most
unstable positions  to
the appearance of new cusps are precisely the midpoints between existing
cusps. This is
the mechanism for the addition of cusps in analogy with tip splitting in
Laplacian growth.

We can now estimate the width of the flame front as the height of the
largest cusps. Since
this height is proportional to $\cal L$ (cf. property (v) of the TFH solution),
Eq.(\ref{scaling1}) and Eq.(\ref{scaling}) lead to the scaling relation
\begin{equation}
\chi = 1/\beta \ . \label{scalrel}
\end{equation}
This scaling law is expected to hold all the way to $\beta=1$ for which the
flame front
does not accelerate and the size of the cusps becomes proportional to $r_0$.

Next we shed light on phenomenon of tip splitting that here is seen as the
addition
of new cusps roughly in between existing ones.  We mentioned the
instability toward the
addition of
new poles. We argue now that the tip between the cusps is most sensitive to
pole creation.
This can be shown in both channel and radial geometry. For example
consider a TFH-giant cusp
solution in which all the poles are aligned (without loss of generality) on
the $x=0$
line. Add a new pole in the complex position $(x_a,y_a)$
to the existing $N(L)$ poles, and study its
fate. It can be shown that in the limit $y_a \to \infty$ (which is the
limit of a
vanishing perturbation of the solution)
the equation of motion is
\begin{equation}
{dy_a \over d\tau} = {2\pi\nu \over L}(2N(L)+1)-1   \quad\quad y_a \to \infty \ .
\label{yinf1}
\end{equation}
Since $N(L)$ satisfies (\ref{NofL}) this equation can be rewritten as
\begin{equation}
{dy_a \over d\tau} = {4\pi\nu \over L}(1-\alpha)    \quad\quad y_a \to \infty \ ,
\label{yinf2}
\end{equation}
where $\alpha=( L/(2\pi\nu)+1)/2 -N(L)$. Obviously $\alpha \le 1$ and it is
precisely
$1$ only when $L$ is $L=(2n+1)2\pi \nu$.  Next it can be shown that
for $y_a$ much larger than $y_{N(L)}$ but not infinite the following is true:
\begin{eqnarray}
{dy_a \over d\tau} &>& \lim_{y_a\to \infty} {dy_a \over d\tau} \quad x_a=0 \\
 {dy_a \over d\tau} &<& \lim_{y_a\to \infty} {dy_a \over d\tau} \quad x_a=\pi
\end{eqnarray}
We learn from these results that there exist values of $L$ for which a pole
that is
added at infinity will have marginal attraction ($dy_a/dt=0$).
Similar understanding can be obtained from a standard stability analysis
without using
pole decomposition. Perturbing a TFH-cusp solution we find linear equations
whose
eigenvalues $\lambda_i$ can be obtained by standard numerical techniques.
Fig.2 presents $Re(\lambda_i)$ as a function of systems size, and shows
very clearly that
(i) all $Re(\lambda_i)$ are non-positive.
(ii) at the isolated values of $L$ for which $L=(2n+1)2\pi\nu$
$Re(\lambda_1)$ and
$Re(\lambda_2)$ become
zero (note that due to the logarithmic scale the zero is not evident)
(iii) There exists a general tendency of all $Re(\lambda_i)$ to approach
zero from
below as $L$ increases. This indicates a growing sensitivity to noise when
the system
size increases. (iv) There exists a Goldsone mode $\lambda_0=0$ due to
translational
invariance.

The upshot of this discussion is that finite perturbations
(i.e. poles at finite $y_a$) will grow if the $x$ position of the pole is
sufficiently
near the tip. The position $x=\pi$ (the tip of the finger) is the most
unstable one.
In the channel geometry this means that noise results in the appearance
of new cusps at the tip of the fingers, but due to the attraction to the
giant cusp  they
move toward $x=0$ and disappear in the giant cusp. In fact, one sees in
numerical
simulations a train of small
cusps   that move toward the giant cusp. Analysis shows that at the same time
the furthest
pole at $y_{N(L)}$ is pushed towards infinity. Also in cylindrical geometry
the most sensitive position to the appearance of new cusps is right between
two existing
cusps independently if the system is marginal (the total number of poles
fits the radius)
or unstable (total number of poles is too small at a given radius). Whether
or not the
addition of a new pole results in tip splitting depends on their $x$ position.
When the distance from existing cusps is larger
than ${\cal L}$ the new poles that are generated by noise will
remain near the tip between the two cusps and will cause tip splitting.

Lastly, we note that without the noisy generation of new poles acceleration
is impossible.
This is seen directly from Eq.(\ref{r0pole}) in which all the terms on the
RHS but unity
go to zero with $r_0\to \infty$, and  the velocity saturates. We need the
noisy appearance of new poles to achieve  acceleration. The precise
connection between the
noise amplitude, system size and acceleration that leads to the computation
of the
exponent $\beta$ is beyond the scope of this letter and will be discussed in a
forthcoming publication.

 \begin{figure}
 \epsfxsize=6truecm
 \epsfbox{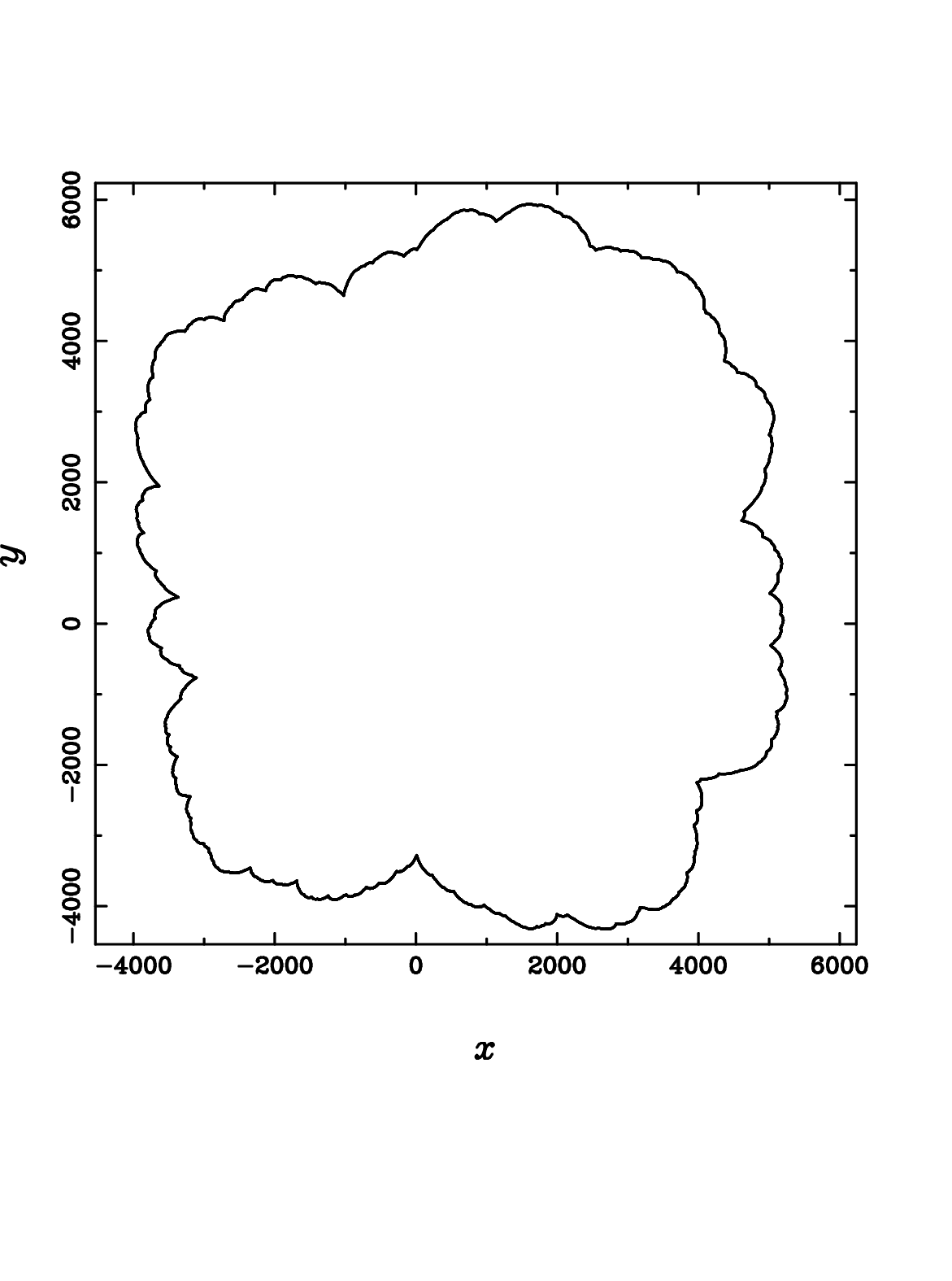}
 \vspace{.5cm}
\vfill
         \hspace{-1.7cm}
 \caption{Simulations of the outward propagating flame front. Note that
deep cusps do
not disappear and that new
deep cusps appear when the rounded tips split.}
\label{fig:fig1}
 \end{figure}

\begin{figure}
 \epsfxsize=6truecm
\epsfbox{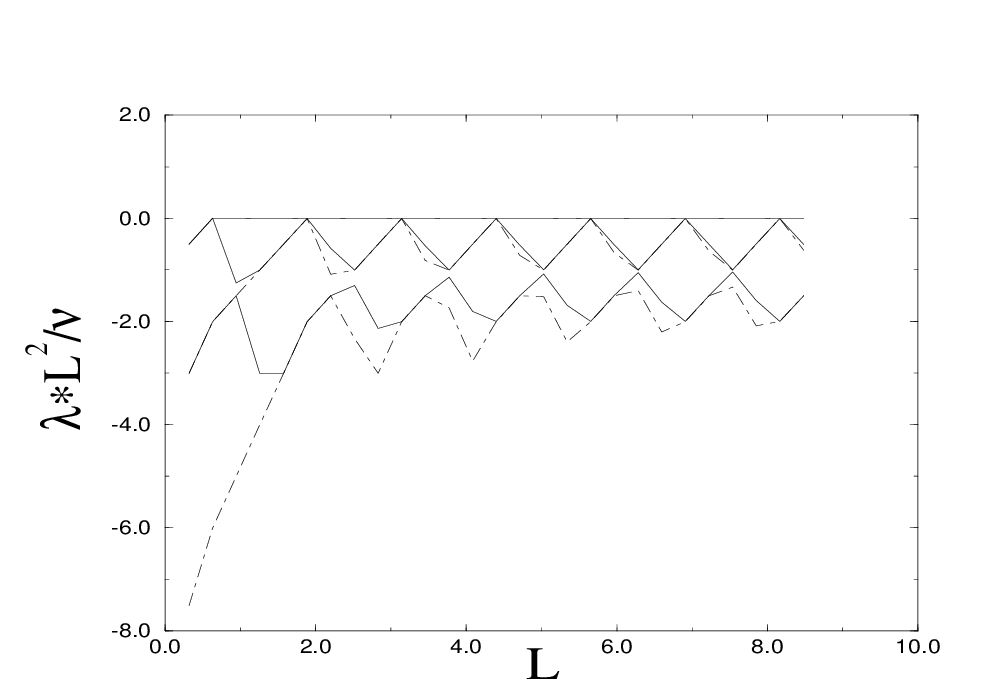}
 \vspace{.5cm}
\caption{Successive eigenvalues of the stability matrix of the
TFH-giant cusp solution as a function of the system size $L$. The
leading eigenvalue touches zero periodically in $L$. All the
eigenvalues tend to zero when $L\to \infty$ as $L^{-2}$.}
\label{fig:fig2}
\end{figure}
\end{document}